\documentclass[twocolumn,showpacs,prl,amsmath,amssymb,floatfix]{revtex4}
\usepackage{dcolumn}
\usepackage{amsmath,amssymb}
\usepackage{bm}
\usepackage{epsfig}

\begin{document}

 \title{Self-energy and critical temperature of weakly interacting bosons} 
                                   
  \author{Sascha Ledowski, Nils Hasselmann, and Peter Kopietz}

  \affiliation{Institut f\"{u}r Theoretische Physik, Universit\"{a}t
    Frankfurt, Robert-Mayer-Strasse 8, 60054 Frankfurt, Germany}

  \date{November 3, 2003}

  \begin{abstract}
Using the exact  renormalization group we calculate  
the momentum-dependent  self-energy 
$\Sigma ( k  )$ at zero frequency  of weakly interacting bosons
at the critical temperature $T_c$ of Bose-Einstein condensation
in dimensions $ 3 \leq D < 4$.
We obtain the complete crossover function 
interpolating between
the critical regime $ k \ll k_c$, where
$\Sigma ( k  ) \propto k^{2 - \eta}$, and
the short-wavelength regime $ k \gg k_c$, where
$\Sigma ( k ) \propto k^{2 (D-3)} $ in $D> 3$ and
$\Sigma ( k ) \propto \ln (k/k_c) $ in $D=3$.
Our approach yields  the 
crossover scale  $k_c$  
on the same footing with
a reasonable estimate for the critical exponent
$\eta$ in $D=3$. 
From our $\Sigma ( k )$ we find for the interaction-induced
shift of $T_c$ in three dimensions
$ \Delta T_c / T_c \approx  1.23 \; a n^{1/3}$,
where $a$ is the s-wave scattering length and $n$ is the density.

  \end{abstract}




  \maketitle

{\it{Introduction.}}
Triggered by Nobel Prize winning experiments,
the past few years have seen an upsurge
of  activity in the field of Bose-Einstein condensation \cite{Pitaevskii03}.
The  interacting Bose gas
falls into the universality class of the classical $O(2)$ model. 
While for dimensions $D$ slightly below the upper critical dimension $D=4$ the small
parameter $\epsilon = 4-D$ can be used to study the
critical behavior via
controlled perturbative  renormalization group (RG) calculations \cite{ZinnJustin89},
no such small parameter exists 
in the physically relevant case $D=3$.
The problem of Bose-Einstein condensation in a 
three-dimensional Bose gas therefore requires
non-perturbative methods.

One of the fundamental properties of the system
is the energy dispersion $\epsilon_k$ of the elementary excitations.
%
In the vicinity of  the critical temperature $T_c$,
where the correlation length becomes large,
interactions qualitatively modify the energy dispersion at long wavelengths.
Precisely at $T_c$ 
it is well known that $\epsilon_k \propto k^{2 - \eta}$
for {\it{sufficiently small momenta}} $k$.
In $D=3$ high order re-summations of the 
$\epsilon$-expansion \cite{Guida98} and 
recent Monte-Carlo simulations \cite{Hasenbusch99}
 yield the critical exponent
$\eta \approx 0.038$.
On the other hand, when $k$ exceeds some interaction-dependent  crossover scale
$k_c$, self-energy corrections to the non-interacting dispersion
$  \rho_0 k^2$ become negligible.
Here $\rho_0 = \hbar^2 / 2 m$, with some effective mass $m$.
%
%
Let us  describe this crossover
in terms of a dimensionless self-energy function 
 \begin{equation}
  \sigma (x ) = ( \rho_0 k_c^2 )^{-1} [ \Sigma ( k_c x ) - \Sigma ( 0 ) ]
 \label{eq:sigmadef}
 \; ,
 \end{equation}
where $\Sigma ( k )$ is the exact self-energy of the interacting Bose gas at zero frequency.
By construction, the crossover occurs at $x = O ( 1 )$.
Note that the field-theoretical RG effectively sets $k_c \rightarrow \infty$, so that neither
$k_c$ nor   $\sigma (x )$ for $x \gg 1$ can be obtained within that framework. 

Naively one might expect that in the short-wavelength regime
$x \gg 1$ the scaling function can be calculated 
perturbatively. Unfortunately, in $D=3$ one 
encounters infrared  divergences even in this regime.
Baym et al.\cite{Baym01} recently
employed a  simple self-consistent re-summation which cures the
divergences  in $D=3$.
However, their  result $\eta=1/2$ is
far from the above mentioned accepted  value \cite{Guida98,Hasenbusch99}.
Clearly, it would be very useful
 to have a quantitatively accurate  formula for  the  momentum-dependence  of the
self-energy at $T_c$, which
interpolates between the critical and the short-wavelength regime.
As emphasized  in Ref. \cite{Baym01}, the 
shift in the critical temperature $\Delta T_c$ due to interactions is closely related
to the behavior of the scaling function $\sigma (x )$; 
at weak coupling and to leading order in the scattering length,  $\Delta T_c$ in $D=3$ 
is proportional to the crossover scale $k_c$, with a
constant of proportionality that depends on the specific form of
$\sigma (x)$ for $0 \leq x \leq \infty$.

{\it{Exact RG flow equations for interacting bosons.}}
Previously  Bijlsma and Stoof \cite{Bijlsma96} used  RG methods 
to investigate the infrared behavior of the weakly  interacting Bose gas
above and below $T_c$.
However, these authors considered only the constant
parts of the vertices. In particular,
no  attempt was made to
calculate the momentum-dependence of the  self-energy.
Here we study this problem using 
the exact  RG 
in the form proposed by Wetterich \cite{Wetterich93}, 
and by Morris \cite{Morris94}.
In this approach one considers the RG flow of the generating  functional
of the one-particle irreducible vertices   $\Gamma^{(2n)}_{\Lambda}$  as 
an infrared cutoff $\Lambda$ is reduced.
Assuming that the bosons are in the normal state \cite{Pistolesi03}, 
the exact flow equation for the
irreducible two-point vertex is \cite{Kopietz01b,footnote1} 
 \begin{eqnarray}
 \partial_\Lambda \Gamma^{(2)}_{\Lambda} ( K ) & = & 
  \int_{ K^{\prime}}
 \dot{G}_{\Lambda} ( K^{\prime} )
 \Gamma^{(4)}_{\Lambda} ( K , K^{\prime} ; K^{\prime} , K )
 \; ,
 \label{eq:flowGamma2}
 \end{eqnarray}
where for sharp infrared cutoff $\Lambda$
\begin{equation}
 \dot{G}_{\Lambda } ( K ) = \frac{ \delta (  | {\bf{k}} |
  - \Lambda )}{  
 i \omega_{n} - \epsilon_{ {\bf{k}} } 
 + \mu  - \Sigma (0 , i0) 
- \Gamma^{(2)}_{\Lambda} ( K ) }
 \; . 
 \label{eq:GLambda}
\end{equation}
We use the notation $K = ( {\bf{k}} , i \omega_n )$,
 $
 \int_K  =  (\beta V)^{-1} \sum_{ {\bf{k}} , \omega_n }
 $,
where $\beta$ is the inverse temperature, $V$ is the volume, and
$\omega_n = 2 \pi n T$ are bosonic Matsubara frequencies.
The  irreducible four-point vertex 
satisfies 
\begin{widetext}
\begin{eqnarray}
 \partial_\Lambda \Gamma^{(4)}_{\Lambda} ( K_1^{\prime} , K_2^{\prime} ; K_2 , K_1 ) & =&
    \int_{ K}  \dot{G}_{\Lambda} ( K )
 \Gamma^{(6)}_{\Lambda} ( K_1^{\prime} , K_2^{\prime} , K  ; K , K_2 , K_1 )
 \nonumber
 \\
 & & \hspace{-38mm} + \int_K
  \left[  \dot{G}_{\Lambda} ( K )
 G_{\Lambda } ( K^{\prime} ) 
 \Gamma^{(4)}_{\Lambda} ( K_1^{\prime} , K_2^{\prime} ; K^{\prime} , K ) 
 \Gamma^{(4)}_{\Lambda} ( K , K^{\prime} , K_2 , K_1 ) 
 \right]_{K^{\prime} = K_1 + K_2 - K }
 \nonumber
 \\
 & & \hspace{-38mm} + 
 \int_K \left[
\bigl[
\dot{G}_{\Lambda} ( K )
 G_{\Lambda } ( K^{\prime} ) 
 +
 G_{\Lambda} ( K )   \dot{G}_{\Lambda} ( K^{\prime} )
 \bigr]
 \Gamma^{(4)}_{\Lambda} ( K_1^{\prime} ,  K^{\prime} ;  K , K_1 ) 
 \Gamma^{(4)}_{\Lambda} ( K_2^{\prime} , K ; K^{\prime} , K_2 ) 
 \right]_{K^{\prime} =  K_1 - K_1^{\prime} + K }
 \nonumber
 \\
 & & \hspace{-38mm}   +
  \int_K \left[
\bigl[
\dot{G}_{\Lambda} ( K )
 G_{\Lambda } ( K^{\prime} ) 
 +
 G_{\Lambda} ( K )   \dot{G}_{\Lambda} ( K^{\prime} )
 \bigr]
 \Gamma^{(4)}_{\Lambda} ( K_2^{\prime} ,  K^{\prime} ;  K , K_1 ) 
 \Gamma^{(4)}_{\Lambda} ( K_1^{\prime} , K ; K^{\prime} , K_2 ) 
 \right]_{K^{\prime} =  K_1 - K_2^{\prime} + K}
 \; .
 \label{eq:flowGamma4}
 \end{eqnarray}
\end{widetext}
Here the exact propagator ${G}_{\Lambda} ( K )$ is obtained
from Eq. (\ref{eq:GLambda}) by replacing
$ \delta (  | {\bf{k}} |- \Lambda ) \rightarrow \Theta (
| {\bf{k}} |- \Lambda )$.
The flow equation for the  irreducible six-point vertex $\Gamma^{(6)}_{\Lambda}$
on the right-hand side of Eq. (\ref{eq:flowGamma4}) can be found in Ref. \cite{Kopietz01b}.
Because quantum mechanics is irrelevant 
for classical critical phenomena, for our purpose
it is sufficient to retain only the zero frequency part of all vertices and work with an
effective classical theory,
with ultraviolet cutoff given by 
 $\Lambda_0 = {2 \pi}/{\lambda_{\rm th}}$, where
 $\lambda_{\rm th} = { h}/{\sqrt{ 2 \pi m  T}}$ is the thermal de Broglie wavelength. 
In principle, the above flow equations can also be used to
calculate the
parameters of the effective classical model, which
are renormalized by the degrees of 
freedom involving non zero Matsubara frequencies. 
Here we take these  finite renormalizations  implicitly into account
via the initial conditions at scale $\Lambda_0$. 
In particular,  in $D=3$ we write 
the initial value of the four-point vertex as
  $\Gamma^{(4)}_{\Lambda_0} ( 0,0;0,0)  = 16 \pi \rho_0 a$, where
$a$ is the $s$-wave scattering length.

To study the critical regime, it is convenient  to introduce dimensionless
momenta ${\bf{q}} = {\bf{k}}/ \Lambda  $ and 
the rescaled classical propagator
$\tilde{G}_l (  q ) = - ( \rho_0 \Lambda^2 / Z_l )
 G_{\Lambda} ( {\bf{q}} \Lambda, i0 ) = \Theta ( q -1 ) / R_l ( q )$,
which is a function of the  
logarithmic flow parameter $l = - \ln ( \Lambda / \Lambda_0 )$.
Here $R_l ( q ) = Z_l q^2 + \tilde{\Gamma}^{(2)}_l ( q )$, where
 $
 \tilde{\Gamma}^{(2)}_l ( q )  = 
   Z_l ( \rho_0 \Lambda^2 )^{-1}
 \Gamma_{\Lambda}^{(2)} ( \Lambda {\bf{q}} , i 0 )$
is the rescaled two-point vertex, and
 $Z_l = 1 - 
 \partial \tilde{\Gamma}^{(2)}_l ( q ) /{\partial q^2} |_{ q^2 = 0 }$
is the wave-function renormalization factor.
For $n \geq 2$ we define the rescaled higher-order vertices 
 \begin{eqnarray}
 \tilde{\Gamma}^{(2n)}_l (    {\bf{q}}_i  )
 & = & ( K_D T )^{n-1}  \Lambda^{ D (n-1) - 2n }
 \left( {Z_l}/{\rho_0} \right)^n 
 \nonumber
 \\
 & \times  & 
 \Gamma_{\Lambda}^{(2n)} (
    {\bf{k}}_i \rightarrow  \Lambda {\bf{q}}_i  ,  \omega_{n_i}  \rightarrow 0     )
 \; .
 \end{eqnarray}
For later convenience we have inserted the numerical factor
$K_D = \Omega_D / ( 2 \pi )^D$, where $\Omega_D$ is the
surface area of the $D$-dimensional unit sphere.
The rescaled vertices satisfy exact flow equations of the form
 \begin{eqnarray}
 \partial_l  \tilde{\Gamma}^{(2n)}_l (   {\bf{q}}_i  )
 & = & \Bigl[ 2n - D ( n-1 ) - n \eta_l 
 \nonumber
 \\
 & & \hspace{-25mm}
 - \sum_{i=1}^n \left(
 {\bf{q}}_i^{\prime} \cdot \nabla_{ {\bf{q}}_i^{\prime}}  + 
 {\bf{q}}_i \cdot \nabla_{ {\bf{q}}_i} \right) 
 \Bigr]  \tilde{\Gamma}^{(2n)}_l (   {\bf{q}}_i  )
  +  \dot{{\Gamma}}^{(2n)}_l (   {\bf{q}}_i  )
 \; ,
 \label{eq:rescaledflowgeneral}
 \end{eqnarray}
where
 $\eta_l  =    \partial \dot{\Gamma}^{(2)}_l ( q ) /
 \partial q^2 |_{ q =0}$
is the flowing anomalous dimension.
The function  $\dot{\Gamma}_l^{(2)} ( q )$
on the right-hand side of the flow function of the 
two-point vertex is
 \begin{eqnarray}
 \dot{\Gamma}_l^{(2)} ( q )
 & = & 
 \int_{\bf{q}^{\prime}}   
 \dot{G}_l ( q^{\prime} ) 
\tilde{\Gamma}^{(4)}_l ( {\bf{q}} , {\bf{q}}^{\prime} ;
 {\bf{q}}^{\prime} , {\bf{q}} )
 \; ,
 \label{eq:gammadot2def}
 \end{eqnarray}
with   
 $\int_{\bf{q}}   =
\int  d^D q /  \Omega_D$
 and
 $\dot{G}_l ( q ) =  \delta ( q -1 )/R_l ( q )$.
The flow function
$\dot{\Gamma}_l^{(4)}
( {\bf{q}}_1^{\prime} , {\bf{q}}_2^{\prime} ; {\bf{q}}_2 , {\bf{q}}_1 )
   $  can be obtained from the
right-hand side of Eq. (\ref{eq:flowGamma4})
by replacing $\dot{G}_{\Lambda} (K) \rightarrow - \dot{G}_l ( q )$,
$G_{\Lambda} (K)  \rightarrow - \tilde{G}_l ( q )$, 
$\Gamma^{(2n)}_{\Lambda}
 \rightarrow \tilde{\Gamma}^{(2n)}_l$, $\int_K \rightarrow 
\int_{\bf{q}}$, and multiplying
the resulting expression by an overall minus sign.
From the above definitions we can show \cite{Ledowski03} that
at $T_c$
the function (\ref{eq:sigmadef}) 
can be written as
 \begin{equation}
 \sigma ( x ) =  \int_0^{\infty} dl
 e^{ - 2 ( l - l_c) + \int_0^{l} d \tau \eta_{\tau} }
  \dot{\Gamma}^{ (2-r)}_{ l } ( e^{ l - l_c } {{x}} )
 \; ,
 \label{eq:sigmaexact}
 \end{equation} 
where $\dot{\Gamma}^{ (2-r)}_{ l } ( q  ) =
\dot{\Gamma}^{ (2 )}_{ l } ( q  ) - \dot{\Gamma}^{ (2 )}_{ l } ( 0  )$ and
$l_c$ will be determined shortly.
Hence, the flow function (\ref{eq:gammadot2def}) is the key quantity for obtaining
$\sigma (x )$ via the RG.

{\it{Classification of couplings and crossover scale.}}
In  $ 3 \leq  D < 4$ there are two  relevant couplings: 
the momentum-independent parts
$r_l = \tilde{\Gamma}_l^{(2)} ( 0 )$ 
and
$u_l = \tilde{\Gamma}_l^{(4)} ( 0 , 0 ; 0 , 0)$
of the
two-point and four-point  vertices. 
If we simply ignore all other couplings, the flow
of $r_l$ and $u_l$ is determined by
 \begin{eqnarray}
 \partial_l r_l & = & ( 2 - \eta_l ) r_l + 
 u_l / ( 1 + r_l ) 
 \label{eq:rlflow2}
 \; ,
 \\
 \partial_l u_l & = & ( \epsilon  - 2 \eta_l ) u_l  - \frac{5}{2} 
 u_l^2 / ( 1 + r_l)^2
 \label{eq:ulflow2}
 \; .
 \end{eqnarray}
Note that the initial value $r_0$ has to be fine tuned
so that $r_{\ast} = \lim_{l \rightarrow \infty} r_l$ is finite,
corresponding to the critical point. 
As long as $ |r_l | \ll 1$ and $\eta_l \ll \epsilon$, the
qualitative behavior of the solution of Eq. (\ref{eq:ulflow2}) can be
obtained by simply replacing $r_l \rightarrow 0$ and $\eta_l
 \rightarrow 0$ on the right-hand side. 
Then the solution of Eq. (\ref{eq:ulflow2}) is
a Fermi function,
 ${u_l}/{u_{\ast} } 
 = [ e^{ \epsilon ( l_c - l ) } +1 ]^{-1}$,
where $u_{\ast} =  2 \epsilon / 5 $ and 
 $l_c = \epsilon^{-1} \ln ( u_{\ast} /{u_0} -1 ) \approx
  \epsilon^{-1} \ln ( u_{\ast} /{u_0} )$. 
Throughout this work we assume
that $u_0 \ll u_{\ast}$, corresponding to weak bare interactions. 
Thus, within a narrow  
interval of width $1/ \epsilon$ centered at $l = l_c$
the coupling $u_l$ suddenly approaches its fixed point value $u_{\ast}$.
Below we shall show that 
$k_c = \Lambda_0 e^{- l_c} \approx 
\Lambda_0 ( u_0 / u_{\ast} )^{1/ \epsilon }$ is  the
scale where the dispersion crosses over from 
$k^2$ to the  $k^{2 - \eta}$.

We now discuss the marginal couplings.
For $3 < D < 4$  the only marginal parameter is  the
wave-function renormalization $Z_l$; it satisfies
by construction $\partial_l Z_l = - \eta_l Z_l$.
In $D = 3$, there are {\it{three}} additional marginal couplings.
The most obvious one \cite{Bijlsma96} is the
momentum-independent part of the six-point vertex,
 ${v}_l = \tilde{\Gamma}^{(6)}_l (0,0,0 ; 0,0,0 )$.
But there are two more marginal
couplings $a_l$ and $b_l$ related to the four-point vertex, which
has the expansion for small momenta
 \begin{eqnarray}
 \tilde{\Gamma}_{l}^{ (4)}
 ( {\bf{q}}_1^{\prime} , {\bf{q}}_2^{\prime} ; {\bf{q}}_2 , {\bf{q}}_1 ) 
 & = & u_l  + {a}_l 
 \left( | {\bf{q}}_1 - {\bf{q}}_1^{\prime} |+  | {\bf{q}}_1 - {\bf{q}}_2^{\prime} |
 \right)
 \nonumber
 \\
 &   + & b_l | {\bf{q}}_1 + {\bf{q}}_2 | + O ( {\bf{q}}_i^2 )
 \; . 
\end{eqnarray}
One easily verifies that in $D=3$ the scaling dimensions of
$a_l$ and $b_l$ vanish. Note that
even if initially
 $\tilde{\Gamma}_{l=0}^{ (4)} = u_0 $
 is momentum-independent (corresponding to the initial conditions
${a}_0 = {b}_0 =0$), finite values of $a_l$ and $b_l$ are
generated as we iterate the RG.

{\it{Truncation and anomalous dimension.}}
To calculate $\eta_l$,  we need the momentum-dependent part of the
four-point vertex. Note that in $D > 3$ the momentum-dependence
of  $\tilde{\Gamma}_{l}^{ (4)}
 ( {\bf{q}}_1^{\prime} , {\bf{q}}_2^{\prime} ; {\bf{q}}_2 , {\bf{q}}_1 ) $
is irrelevant. As explained in Refs. \cite{Polchinski84,Kopietz01}, 
for sufficiently large $l$ irrelevant couplings
become local functions of the relevant and marginal ones.
Assuming $u_l \ll 1$, 
the momentum-dependence  of
 $\tilde{\Gamma}_{l}^{ (4)}
 ( {\bf{q}}_1^{\prime} , {\bf{q}}_2^{\prime} ; {\bf{q}}_2 , {\bf{q}}_1 ) $
can be obtained by setting 
$\tilde{\Gamma}^{(6)}_{l} \rightarrow 0$ and
$ \tilde{\Gamma}_{l}^{(4) } \rightarrow u_l$
in the expression for the flow function
 $\dot{\Gamma}_{l}^{(4) }
( {\bf{q}}_1^{\prime} , {\bf{q}}_2^{\prime} ; {\bf{q}}_2 , {\bf{q}}_1 )$.
In Ref. \cite{Busche02} we have used a similar truncation to calculate
the spectral function of the Tomonaga-Luttinger model using the exact RG.
In the present problem this truncation amounts to 
 \begin{eqnarray}
 \dot{\Gamma}_{l}^{(4) }
( {\bf{q}}_1^{\prime} , {\bf{q}}_2^{\prime} ; {\bf{q}}_2 , {\bf{q}}_1 )
 & \approx &   - {u}_l^2 
 \Bigl[ \frac{1}{2} \dot{\chi}_l ( |  {\bf{q}}_1 + {\bf{q}}_2 | )
 \nonumber
 \\
 & & \hspace{-20mm}
 + \dot{\chi}_l (| {\bf{q}}_1 - {\bf{q}}_{1}^{\prime}  | )
 + \dot{\chi}_l (| {\bf{q}}_1 - {\bf{q}}_{2}^{\prime} | ) 
 \Bigr] 
 \; \;  ,
 \label{eq:dotGamma4approx}
 \end{eqnarray}
where the generalized susceptibility is given by
 $\dot{\chi}_l ( q )  =   2 \int_{ {\bf{q}}^{\prime}}
 \dot{G}_l ( |  {\bf{q}}^{\prime} | )  \tilde{G}_l ( | {\bf{q}}^{\prime} + {\bf{q}}|  )$.
The approximation (\ref{eq:dotGamma4approx})
can be formally justified as long as
$u_l$ remains small, which
is certainly the case for small $\epsilon$.
Unfortunately, in $D=3$
the renormalized $u_{\ast}$ is 
not guaranteed to be small.
Moreover, the  couplings
$v_l$, $a_l$ and $b_l$ which become marginal in $D=3$ are not consistently
taken into account in Eq. (\ref{eq:dotGamma4approx}), so that {\it{a priori}} 
we cannot estimate the accuracy of our truncation in $D=3$.
Nevertheless, we shall use
Eq. (\ref{eq:dotGamma4approx}) as a first approximation
even in $D=3$; an improved truncation including
the marginal couplings $v_l$, $a_l$ and $b_l$ will be presented elsewhere \cite{Ledowski03}.

Given the truncation (\ref{eq:dotGamma4approx}), the exact flow equation 
(\ref{eq:rescaledflowgeneral}) for $
 \tilde{\Gamma}_{l}^{(4) }
( {\bf{q}}_1^{\prime} , {\bf{q}}_2^{\prime} ; {\bf{q}}_2 , {\bf{q}}_1 )$
can be solved without further approximation.
Substituting the result into Eq. (\ref{eq:gammadot2def}) we obtain for
 the function $\dot{\Gamma}^{ (2-r)}_{ l } ( q  )$
required in Eq. (\ref{eq:sigmaexact})
 \begin{eqnarray}
 \dot{\Gamma}_l^{(2-r)} ( {{q}} ) & \approx &
  - \frac{3}{2(1 + r_l)} 
 \int_0^{l} d t e^{  \epsilon t   -  2 \int_{l-t}^l d \tau \eta_{\tau}
 } {u}_{l-t}^2
 \nonumber
 \\ & & \hspace{-11mm} \times
 \langle 
 \dot{\chi}^{}_{l-t} (  e^{ - t}  |  \hat{\bf{q}}^{\prime} + {\bf{q}} |  )
 -  \dot{\chi}^{}_{l-t} (  e^{ - t}  ) 
 \rangle_{\hat{\bf{q}}^{\prime} }
 \; ,
 \label{eq:gamma2miapprox}
 \end{eqnarray}
where $\langle \ldots \rangle_{ \hat{\bf{q}}^{\prime} }$ denotes angular average over the
unit vector $\hat{\bf{q}}^{\prime}$.
Eq. (\ref{eq:gamma2miapprox}) implies the following 
integral equation for the flowing anomalous dimension
 $\eta_l  =    \partial \dot{\Gamma}^{(2)}_l ( q ) /
 \partial q^2 |_{ q =0}$,
 \begin{equation}
  \eta_l =    \int_0^{l} d t K (l,t)  u_{l-t}^2
 e^{  - 2 \int_{l-t}^l d \tau \eta_{\tau} }
 \; ,
 \label{eq:etaintegral}
 \end{equation}
with the kernel given by
 \begin{eqnarray}
 K ( l , t )  
 & = & - \frac{3}{4 D (1 + r_l)} \Bigl[ (D-1 ) e^{- ( D-3 ) t }
 \dot{\chi}^{\prime}_{l-t} ( e^{-t} ) 
 \nonumber
 \\
 & & \hspace{10mm} + e^{- ( D-2) t }
  \dot{\chi}^{\prime \prime}_{l-t} ( e^{-t} ) \Bigr]
 \; ,
 \label{eq:Kltdef}
 \end{eqnarray}
where $ \dot{\chi}^{\prime}_{l} ( x ) = d \dot{\chi}_l ( x ) / d x $ and
$ \dot{\chi}^{\prime \prime}_{l} ( x ) = d^2 \dot{\chi}_l ( x ) / d x^2 $.
Eq. (\ref{eq:etaintegral}) together with the two
flow equations (\ref{eq:rlflow2},\ref{eq:ulflow2}) form a system of three coupled
integro-differential equations for the three unknown functions
$r_l$, $u_l$ and $\eta_l$.  For small $\epsilon$ we recover the 
known result $\eta = \lim_{ l \rightarrow \infty} \eta_l = \epsilon^2 / 50$.
In $D=3$ we find numerically at the fixed point
$r_{\ast} = -0.143$, $u_{\ast} =
0.232$, and $\eta = 0.104$, which should be compared with 
the  accepted value $\eta = 0.038$ \cite{ZinnJustin89,Guida98,Hasenbusch99}.
Given the simplicity of our truncation,
it is quite satisfactory that our estimate for  $\eta$ has  
the correct order of magnitude.
We can substantially improve our estimate for $\eta$ 
by including the above mentioned marginal couplings \cite{footnoteimprove}.
In comparison,  the method  of Ref. \cite{Baym01}
yields $\eta = 0.5 $.

\begin{figure}[tb]
\epsfysize6cm 
\hspace{5mm}
 \epsfig{file=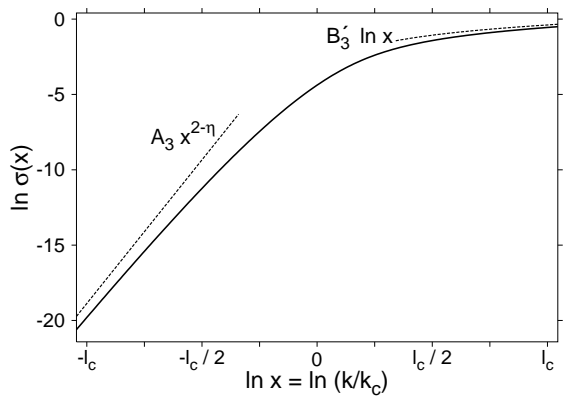,width=75mm}
\vspace{5mm}
\caption{
Double logarithmic plot of the scaling function $\sigma (x )$
in $D=3$, see Eqs.  (\ref{eq:sigmadef}) and (\ref{eq:sigmares}). 
Note that in our  rescaled units we obtain a universal  curve
for $u_0 \rightarrow 0$.
The dashed lines indicate the asymptotic behavior for
small and large $x$.
}
\label{fig:sigma}
\end{figure}

{\it{Scaling function.}}
To obtain the
scaling function $\sigma (x )$ we substitute
our approximation (\ref{eq:gamma2miapprox}) into
Eq.(\ref{eq:sigmaexact}). 
The integrations can be simplified
by noting that the flow of $\eta_{\tau}$
on the right-hand side can be approximated by a Fermi function,
 $\eta_\tau  = \eta [ e^{ \epsilon ( l_c - \tau ) } +1 ]^{-1}$.
We have verified numerically that an analogous substitution  
in Eq. (\ref{eq:etaintegral})
yields a good approximation to the true solution $\eta_l$.
Introducing appropriate integration variables we obtain 
 \begin{equation}
 \sigma ( x ) = \frac{3 u_{\ast}^2}{2} x^{ 2 - \eta} 
 \int_{ x e^{-l_c}  }^{\infty} d y
 \frac{ y^{ -3 + 2 \epsilon}  F ( x, y; \eta , l_c)    }{[ x^{\epsilon} + y^{\epsilon} ]^{ 2 - 2 \eta / \epsilon} }
 \; ,
 \label{eq:sigmares}
 \end{equation}
with the dimensionless function
 \begin{eqnarray}
 F (  x , y ;\eta, l_c)
 & = & \int_{0}^{1} d z \frac{ z^{1 - \epsilon}}{ \left[ x^{\epsilon} + ( y/z )^{\epsilon} 
 \right]^{ \eta / \epsilon} \bigl[
 1 + r_{l_c + \ln \frac{y}{zx} } \bigr] }
 \nonumber
 \\
 &  & \hspace{-13mm} \times 
\langle \dot{\chi}_{ l_c + \ln \frac{ y}{x} } ( z  ) 
 - \dot{\chi}_{ l_c + \ln \frac{y}{x} } ( | z \hat{\bf{q}}^{\prime} + y  \hat{\bf{q}}  | )
\rangle_{ \hat{\bf{q}}^{\prime} }
 \; .
 \label{eq:FDdef}
 \end{eqnarray}
A numerical evaluation of  Eq. (\ref{eq:sigmares}) is shown in 
Fig.~\ref{fig:sigma}. 
The asymptotic behavior for small and large $x$ 
can be extracted analytically. In the critical regime $x \ll 1$
we find in $D$ dimensions
 $\sigma (x ) \sim A_D x^{ 2 - \eta }$, with
 $A_D  =  
\frac{3}{2} u_{\ast}^2 
 \int_{ 0  }^{\infty} d y y^{ -3 + 2 \eta} 
  F ( 0 , y; \eta , \infty)$.
In $D=3$ we obtain numerically $A_3 = 1.20$.
On the other hand, in the short wavelength regime 
we find
 $\sigma (x ) \sim B_D x^{ 2  ( D-3) }$
 for  $ [ 2 (D-3)]^{-1}  \ll \ln x$,
and  $\sigma (x ) \sim B_3^{\prime} \ln x $
for $1 \ll \ln x \ll [ 2 (D-3)]^{-1}$ 
(which includes the limit $D \rightarrow 3$).
Here  
$B_D \propto u_{\ast}^2 /(D-3)$ and $B_3^{\prime} = 3 \pi^2 u_{\ast}^2 /24$. 
Keeping in mind that $k_c \propto ( u_0 / u_{\ast})^{1/ \epsilon}$, it is easy to see that 
for  $k \gg k_c$
the physical self-energy $\Sigma ( k ) - \Sigma (0)$ is proportional to
$u_0^2$.

{\it{Critical temperature.}}
Given $\sigma (x )$, we may calculate 
the interaction-induced shift $\Delta T_c$ of the critical
temperature  of the weakly interacting Bose gas.
Generalizing an expression given by Baym et al. \cite{Baym01}
for arbitrary $D$, we obtain for the
contribution from classical fluctuations
  \begin{eqnarray}
 \frac{ \Delta T_c}{T_c}
 & = & 
\frac{2 \Omega_D   ( u_0 / u_{\ast})^{ \frac{ D-2}{4-D} }    }{\pi D \zeta ( D/2 )}  
 \int_0^{\Lambda_0 / k_c} \hspace{-3mm} d x 
 \frac{ x^{D-3}  \sigma ( x ) }
{ x^2 +  \sigma ( x ) } \; .
 \label{eq:tcshift}
 \end{eqnarray}
Keeping in mind that 
$\sigma ( x ) \sim B_D x^{ 2 ( D-3)}$ for $D>3$ and large $x$,
it is easy to see that 
we may remove the ultraviolet cutoff ($\Lambda_0 \rightarrow \infty$)
in Eq.(\ref{eq:tcshift}) provided
 $ \frac{ D-2}{4-D} < 2$, i.e. $D < 10/3$. 
Only in this case
 $\Delta T_c$ is dominated by classical fluctuations.
For 
$D \geq  10/3$ the value of the integral in Eq. (\ref{eq:tcshift}) depends
on the ultraviolet cutoff, such that
it  is proportional to $ u_0^2$, with logarithmic corrections
($ \propto u_0^2 \ln u_0$)
in $D = 10/3$. In three dimensions, where
$u_0 = 16 \pi^{-1} [ \zeta (3/2) ]^{-1/3} a n^{1/3}$, 
we find \cite{footnote2} to leading order
 $\Delta T_c   \approx  c_1  a n^{1/3} T_c$ with
$c_1 = 1.23$,
in agreement with the
five-loop result $c_1 = 1.14 \pm 0.11$ by
Kleinert \cite{Kleinert03} and the seven-loop result
$c_1 = 1.27 \pm 0.11$ by Kastening \cite{Kastening03}.
For recent reviews on the problem of calculating $\Delta T_c$ 
see Ref. \cite{Haque03}.

{\it{Summary and outlook.}}  
In this work we have used the exact RG to calculate the momentum-dependent
self-energy  of weakly interacting bosons at the
critical point.
Our central result is an explicit expression for the scaling function
$\sigma ( k / k_c )$ (see Eqs. (\ref{eq:sigmadef}) and (\ref{eq:sigmares})) which interpolates between the
critical regime $k \ll k_c$ and the short-wavelength regime $ k \gg k_c$.
Note that  the momentum-dependent part
of Eq. (\ref{eq:dotGamma4approx})
contains an infinite set of  irrelevant (and hence non-renormalizable) couplings, so
that neither the crossover scale $k_c$ nor $\Sigma ( k )$ 
for $k \gtrsim k_c$ are accessible within field-theoretical RG.
The crossover  function shown in
Fig.~\ref{fig:sigma}  should be useful for a comparison with experiments probing the
single-particle excitations in a wide range of length scales, for example via
measurements of the velocity distribution.
We have shown here how the exact RG can be employed to calculate full
scaling functions and not just their asymptotic behavior.
Our method is quite general and should  prove useful in the
study of other critical systems, for example Luttinger liquids \cite{Busche02} 
or systems described by the dilute Bose gas quantum critical point \cite{Sachdev00}.


This work was partially supported by the DFG via Forschergruppe FOR 412.


\begin{thebibliography}{99}
\bibitem{footnoteimprove}
Retaining apart from the relevant and marginal parameters $r_l,u_l, v_l, a_l$, and  $b_l$  
also the  irrelevant couplings proportional to $u_l^2$ 
contained in Eq. (\ref{eq:dotGamma4approx}),  we obtain
$\eta = 0.034$ in $D=3$ \cite{Ledowski03}, which is
quite close to the accepted value $\eta = 0.038$.
\bibitem{footnote2}
Since we are interested in the leading term in powers of
$a$, we may simplify the integrand in Eq. (\ref{eq:FDdef}) by
approximating $r_l \approx r_{\ast} [ e^{ l_c - l} +1 ]^{-1}$, which is quantitatively accurate
for $l \gtrsim l_c$ and leads to negligible corrections for $l \lesssim l_c$.
If we simply set $r_l \rightarrow 0$ in Eq. (\ref{eq:FDdef}) we obtain
$c_1 \approx 0.928$, while the replacement
 $r_l \rightarrow r_{\ast}$  yields  $c_1 \approx 1.52$.
\end{thebibliography}
\end{document}